\documentclass[aps,twocolumn,prl,nofootinbib,amsmath,amssymb,amsfonts,superscriptaddress,notitlepage]{revtex4-2}
\usepackage[varg]{txfonts}

\usepackage[T1]{fontenc}
\usepackage[utf8]{inputenc}
\usepackage{hyperref}
\usepackage{amsmath}
\usepackage{color}
\usepackage{float}
\usepackage{graphicx}
\usepackage[percent]{overpic}
\usepackage{mathrsfs}
\usepackage{bm}
\usepackage{braket}
\usepackage{mathtools}

\definecolor{cL}{RGB}{59, 83, 140}
\definecolor{cM}{RGB}{33, 145, 141}
\definecolor{cH}{RGB}{95, 202, 98}
\definecolor{cG}{RGB}{204,204,204}

\newcommand{\bs}[1]{\boldsymbol{#1}}

\def\be{\begin{equation}}
\def\ee{\end{equation}}

\definecolor{cL}{RGB}{59, 83, 140}
\definecolor{cM}{RGB}{33, 145, 141}
\definecolor{cH}{RGB}{95, 202, 98}
\definecolor{cG}{RGB}{204,204,204}

\begin{document}

\title{Magnon Interference Tunneling Spectroscopy as a Probe of 2D Magnetism}

\author{Asimpunya~Mitra}
\affiliation{Indian Institute of Technology Kharagpur, Kharagpur 721302, India}
\author{Alberto Corticelli}
\affiliation{Max Planck Institute for the Physics of Complex Systems, N\"{o}thnitzer Str. 38, 01187 Dresden, Germany}
\author{Pedro Ribeiro}
\affiliation{CeFEMA, Instituto Superior T\'{e}cnico, Universidade de Lisboa, Av. Rovisco Pais, 1049-001 Lisboa, Portugal}
\affiliation{Beijing Computational Science Research Center, Beijing 100084, China}

\author{Paul~A.~McClarty}
\affiliation{Max Planck Institute for the Physics of Complex Systems, N\"{o}thnitzer Str. 38, 01187 Dresden, Germany}

\begin{abstract}
Probing two-dimensional single-layer quantum magnets remains a  significant challenge. 
In this letter, we propose exploiting tunneling spectroscopy in the presence of magnetic impurities to obtain information about the magnon dispersion relations in analogy to quasiparticle interference in non-magnetic materials.
We show this technique can be used to establish the dispersion relations even for frustrated magnets, where the presence of an impurity generally leads to a nontrivial spin texture. 
Finally, we consider the problem of establishing Chern magnon bands in 2D magnets showing how tuneable impurities allow probing the nature of the surface states. 

\end{abstract}

\maketitle

%%%%%%%%%%%%%%%%%%%%%%%%%%%%%%%%%%%%%%%%%%%%%%%%%%%%%%%%%%%%%%%%%%%%%%%%%%%%%%%%%%%%%%%%%%%
%%% INTRODUCTION
%%%%%%%%%%%%%%%%%%%%%%%%%%%%%%%%%%%%%%%%%%%%%%%%%%%%%%%%%%%%%%%%%%%%%%%%%%%%%%%%%%%%%%%%%%%

One of the most exciting recent developments in magnetism is the fabrication of truly 2D magnets on nonmagnetic substrates opening up the possibility of both engineering and exploring novel exotic and interesting low dimensional collective phenomena. Notable examples include twisted bilayer graphene, that under certain circumstances becomes insulating with spontaneous ferromagnetism and Chern bands \cite{Cao2018,Sharpe2019,Serlin2020}, and the chromium trihalides that are insulating honeycomb ferromagnets \cite{Huang2017CrI3,Burch2018,Cai2019,Gibertini2019}. With advances in materials science, there appear to be excellent prospects for the proliferation of engineered few layer magnetic materials with many different kinds of magnetic ion arranged on various lattices and substrates.

A pressing challenge in this area is to develop new techniques to probe such systems. Traditional techniques must be replaced with more microscopic probes. Examples include magnetometry using nitrogen vacancy centres $-$ that has potential as a probe of magnetic structures and low energy magnetic dynamics \cite{Casola2018} $-$ and using Hall currents for the uniform susceptibility \cite{Kim2019}. In this paper, we propose using impurity-induced magnon interference tunneling spectroscopy to probe the magnetic properties of 2D materials, as a solution to the long-standing problem of characterising magnetism in these materials.

Scanning tunneling microscopy (STM) is a well-established technique to probe local charge and spin properties. In recent landmark work, the viability of STM to observe magnetic excitations has also been successfully demonstrated  \cite{Klein2018,Ghazaryan2018}. In many electronic materials, it is possible to infer non-local charge properties by measuring the spatial response to isolated impurities. This technique, called quasi-particle interference (QPI), has also been used to obtain crucial information about electronic correlations \cite{Davis1999,Yazdani1999,Balatsky2006}. 

%%%%%%%%%%%%%%%%%%%%%%%%%%%%%%%%%%%%%%%%%%%%%%%%%%%%%%%%%%%%%%%%%%%%%%%%%%%%%%%%%%%%%%%%%%%
%%% FIGURE: CONDUCTANCE VS VOLTAGE
%%%%%%%%%%%%%%%%%%%%%%%%%%%%%%%%%%%%%%%%%%%%%%%%%%%%%%%%%%%%%%%%%%%%%%%%%%%%%%%%%%%%%%%%%%%
\begin{figure}
  \centering
  \includegraphics[width=\columnwidth]{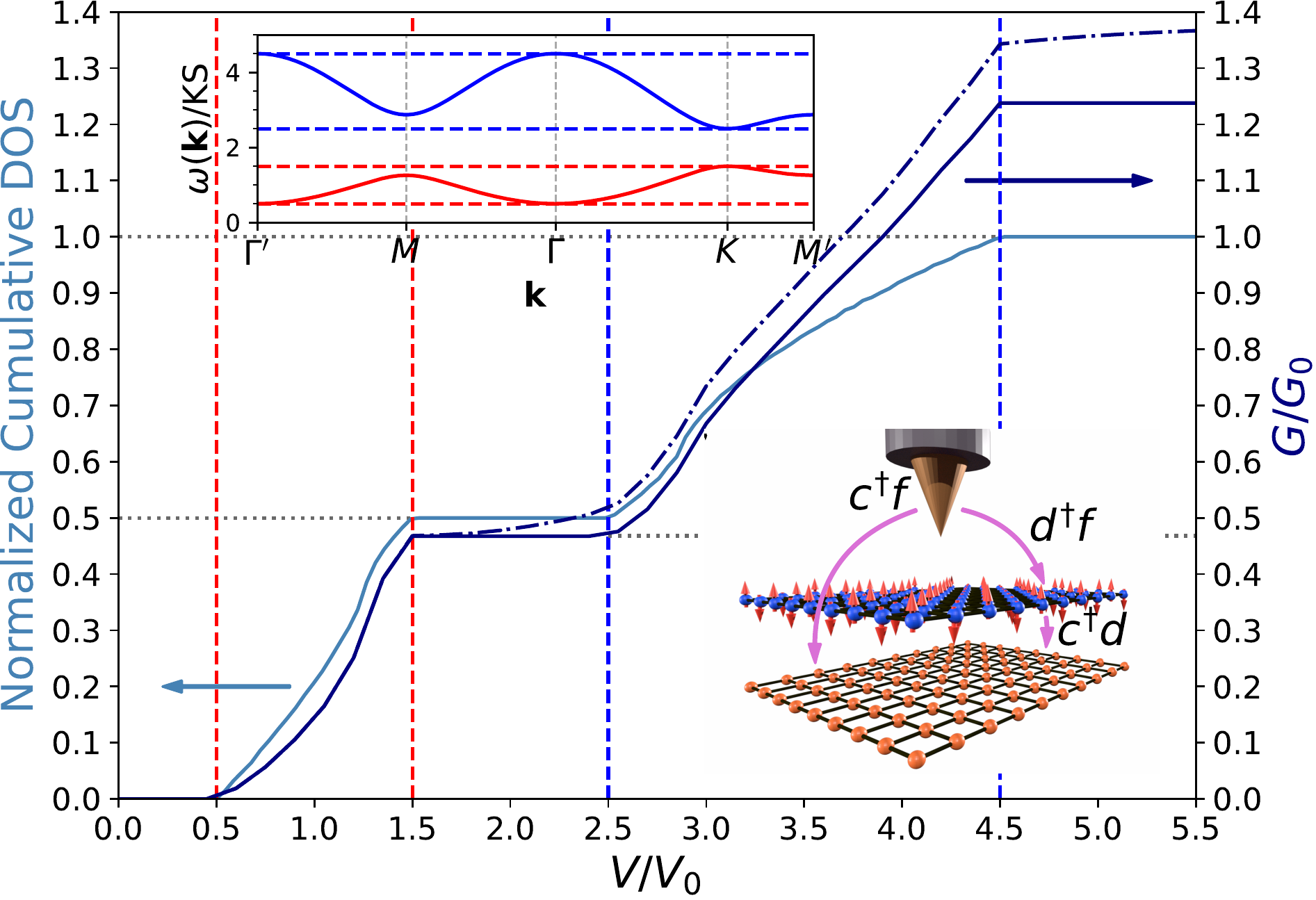}
  \caption{\label{fig:G_No_Impurity} Normalized cumulative density of states (left axis) and normalized inelastic tunneling conductance $G/G_0$ (right axis) plotted as a function of bias voltage $V$ for the Kitaev-Heisenberg model with $\theta=\pi/2$ and $h=4.5$ and the STM with $a/\lambda=1$. The dashed line shows the inelastic conductance with both single magnon and 2-magnon continuum contributions taken into account. The top inset shows the dispersion relations along the indicated high symmetry directions in reciprocal space, dashed lines indicate the band edges. The bottom inset shows the schematic of an STM setup, arrows show the various tunneling processes. 
  }
\label{fig:G_No_Impurity}
\end{figure}

Here, we show that the magnetic analog of QPI - magnon interference tunneling spectroscopy - can be used to study isolated magnetic layers on conducting substrates through the magnetic scattering of tunneling electrons by impurities. In particular, this technique can provide momentum and energy-resolved information about magnons in ordered two-dimensional magnets.

The paper is organised as follows. First, we obtain the STM conductance of electrons tunneling from an STM tip into the conducting substrate that crucially includes the inelastically scattering from magnons in the two-dimensional magnet. The leading contribution to the conductance is shown to depend on the local dynamical structure factor (LDSF) by considering various models with different degrees of approximation. 

We then turn to the modelling of an impurity in an otherwise undisturbed 2D lattice of magnetic ions and study its STM interference pattern in real and momentum space. We illustrate how interference effects can be used to infer magnon dispersion relations in various cases, including (i) single band unfrustrated magnets, (ii) multiple band unfrustrated magnets with touching points (iii) frustrated magnets with multiple bands, where the magnetic ground state itself is destabilized in the vicinity of the impurity \cite{Villain1979}. Finally, we show the potential of engineered impurity scattering to probe properties of Chern magnon bands.

%%%%%%%%%%%%%%%%%%%%%%%%%%%%%%%%%%%%%%%%%%%%%%%%%%%%%%%%%%%%%%%%%%%%%%%%%%%%%%%%%%%%%%%%%%%
%%% SECTION: TUNNELING THROUGH SINGLE LAYER
%%%%%%%%%%%%%%%%%%%%%%%%%%%%%%%%%%%%%%%%%%%%%%%%%%%%%%%%%%%%%%%%%%%%%%%%%%%%%%%%%%%%%%%%%%%

{\it Tunneling of electrons through a magnetic layer} $-$ 
Before introducing any impurity, we consider the setup shown in Fig.~\ref{fig:G_No_Impurity}(a) featuring an STM tip over an insulating magnetic layer mounted on a conducting substrate.  
We begin by calculating the tunneling conductance originating from inelastic scattering from localized moments
when a bias voltage $V$ is applied between the tip and the substrate.
Similar situations were considered in Refs. \cite{Applebaum1966,Anderson1966,Pustilnik2001, Schrieffer1966, Maltseva2009, Fransson2010, Feldmeier2020,Udagawa2021}. Here, we extend their analysis to highlight various points that are peculiar to the case of magnetic scattering through a monolayer such as the role of the extended substrate and the multiboson states.

The Hamiltonian for this system is composed of the tip Hamiltonian $H_{\rm T} = \sum_{\bs{k},\sigma} \epsilon_{{\rm T}\bs{k}} f_{\bs{k}\sigma}^\dagger f_{\bs{k}\sigma}$, the substrate Hamiltonian $H_{\rm S} = \sum_{\bs{k},\sigma} \epsilon_{{\rm S}\bs{k}} c_{\bs{k}\sigma}^\dagger c_{\bs{k}\sigma}$ and the Hamiltonian for the impurity layer $H_{\rm mag} = \sum_{\bs{r}_{\rm I}, \sigma} \epsilon_d d^\dagger_{\bs{r}_{\rm I}} d_{\bs{r}_{\rm I}} + U \sum_{\bs{r}_{\rm I}} n_{d_{\bs{r}_{\rm I} \uparrow}}n_{d_{\bs{r}_{\rm I} \downarrow}}$.
In addition, we consider a tunneling term with amplitude $t_1$ to hop from the tip to a magnetic atom and amplitude $t_2$ to hop to and from the magnetic layer to the substrate.

As the charge gap of the magnetic layer is assumed to be much larger than any other energy scale in the problem, we may derive an effective Hamiltonian for the impurity layer in terms of localized moments to leading order in $1/U$ and $1/\epsilon_d$ \cite{SM}. The resulting Hamiltonian is $\mathcal{H} = \mathcal{H}_0 + \delta \mathcal{H}$, where $\mathcal{H}_0$ is the unperturbed Hamiltonian for the tip and substrate and $\delta \mathcal{H} =   \sum_{r_{\rm I},r_{\rm I}', r_{\rm S}} J(\bs{r}_{\rm T}-\bs{r}_{\rm I},\bs{r}_{\rm I}'-\bs{r}_{\rm S}) \bs{S}_{\bs{r}_{\rm I},\bs{r}_{\rm I}'}\cdot \bs{s}_{\bs{r}_{\rm T},\bs{r}_{\rm S}}+\cdots$, where $S^i_{\bs{r}_{\rm I},\bs{r}_{\rm I}'} = d^\dagger_{\bs{r}_{\rm I},s} \sigma^{i}_{s,s'} d_{\bs{r}_{\rm I}',s'}$ and  $s^{i}_{\bs{r}_{\rm T},\bs{r}_{\rm S}} = f^\dagger_{\bs{r}_{\rm T},s} \sigma^{i}_{s,s'} c_{\bs{r}_{\rm S}, s'}$, 
induces inelastic hopping mediated by the magnetic atoms, terms denoted by $...$ do not contribute to the linear response current.
Therefore, within linear response, the current is given by $I \equiv e dN_{\rm tip}/dt $ \cite{SM}.  

The tip is positioned laterally at $\boldsymbol{r}_{\rm T}$ with vertical distance $d_1$ just above a magnetic layer and vertical distance $d_2$ between the layer and the substrate. We denote lattice constant of the magnetic layer as $a$. In principle, all hopping events, from the tip to any magnetic atom and from a magnetic atom to any position in the substrate, should be taken into account. Henceforth, we assume that the tunneling into the substrate happens only at position $\bs{r}_{\rm T}$. In the Supplementary Material (SM) \cite{SM} we relax this assumption and show that the exponential suppression of the tunneling with distance justifies this approximation. Within this approximation the net tunneling amplitude $T(\bs{r}_{\rm I} - \bs{r}_{\rm T})=t_1^{*}(\bs{r}_{\rm I} - \bs{r}_{\rm T}) t_2(\bs{r}_{\rm I} - \bs{r}_{\rm T}) \simeq \Gamma^2 \exp(-(d_1+d_2)/d) \exp(- \vert \bs{r}_{\rm I} - \bs{r}_{\rm T} \vert /\lambda)$, \cite{SM} decays exponentially with distance where $d$ is a characteristic length scale for tunneling along $z$ direction and $\lambda$ for lateral direction. The $z$ dependence for tunneling is subsequently absorbed into the definition of $G_0$ \cite{SM}.

 \begin{figure*}
  \centering
  \includegraphics[width=2.2\columnwidth,clip,trim={3cm 0 0 0}]{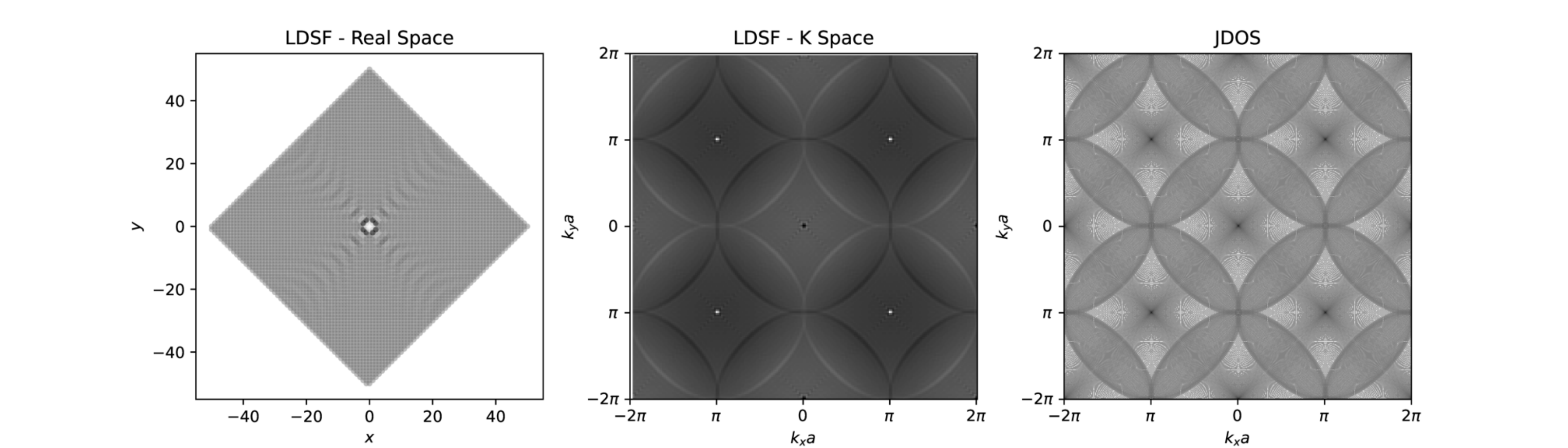}
  \caption{\label{fig:impurity} Figure showing the effect of introducing a single impurity into the square lattice antiferromagnet at energy $\omega/4JS=0.875$. The plot on the left shows the local dynamical structure factor on each site at fixed energy for a $2\times 50\times 50$ site slab computed using the T matrix approach. This is proportional to the differential tunneling conductance at fixed voltage in the limit where tunneling can only occur directly to the site beneath the tip. The middle panel shows the Fourier transformed real space pattern. For comparison we show the joint density of states (right). 
}
\end{figure*}

Finally, we consider the tip and the substrate to be wide-band metals with a constant density of states. This approximation is appropriate in a typical situation where the magnetic energy scales are small as compared to electronic scales.
Under these assumptions, the tunneling conductance is
\begin{equation}
\langle G(V) \rangle = G_0\sum_{ab} \sum_{\bs{r}_{\rm I} \bs{r}_{\rm I}'}e^{-\frac{2}{\lambda}(|\bs{r}_{\rm Ia}-\bs{r}_{\rm T}|+|\bs{r}_{\rm Ib}'-\bs{r}_{\rm T}|)}\int_{0}^{eV}d\omega\,\sum_\alpha\tilde{S}^{\alpha\alpha}_{\bs{r}_{\rm Ia} \bs{r}_{\rm Ib}'}(\omega)
\label{eq:conductance}
\end{equation}
where 
$\sum_i\tilde{S}^{\alpha\alpha}_{\bs{r}_{\rm Ia} \bs{r}_{\rm Ib}'}(\omega)=\frac{1}{2\pi}\int d\mathbf{k}\,\sum_{i}S^{ii}_{ab}(\mathbf{k},\omega)e^{-i\mathbf{k}\cdot(\bs{r}_{\rm Ia}-\bs{r}_{\rm Ib}')}$
$I, I'$ are primitive cell labels and $a,b$ sublattice labels and $S^{\alpha\alpha}_{ab}(\mathbf{k},\omega)$ is the $\alpha$ spin component of the momentum-energy dependent dynamical structure factor (DSF) \cite{SM}. Within linear spin wave theory, the DSF has a single magnon contribution coming from the transverse spin components within the local quantization frame while a two-magnon contribution arises from the longitudinal component \cite{SM,MahanBook}.

Note that in $\mathcal{H}_0$ we include all effective magnetic interactions within the layer that arise from electron-electron interactions in the presence of a small intra-layer hopping. However, the presence of a conducting layer may affect the effective magnetic exchange, for example through RKKY interactions. The effects of the substrate may be even more severe: for example under different conditions the layer may undergo charge disproportionation. Our calculation can, in principle, be adapted to the case where magnetic order co-exists with charge ordering. For illustration, we consider here various exchange models and evaluate the DSF within spin wave theory \cite{SM}.

Figure~\ref{fig:G_No_Impurity}(b) shows the conductance and cumulative density of states as a function of voltage for the honeycomb lattice with Kitaev exchange couplings and with moments polarized perpendicular to the plane by an external magnetic field \cite{McClarty2018}.  
This model has two dispersive magnon bands at finite energy as shown in the figure inset. 
The inelastic component of the conductance is zero up to the voltage corresponding to the base of the lowest magnon band. Then, the tunneling conductance increases as more scattering channels become available at higher energies. The band gap is visible as a plateau in the conductance. The tunneling conductance plateaus again above the maximum energy of the highest magnon band. In short, the on-site tunneling supplies information about the magnon cumulative density of states and reveals the presence of band gaps. Further examples are given in the SM \cite{SM}. 

As one is usually interested more in the density of states than in the cumulative density of states, it is convenient to consider the {\it differential} conductance $G(V+\delta V)-G(V)$  for different on-site positions where $\delta V$ is a current averaging window. 

The calculations discussed so far are based on single magnon inelastic scattering. However, the spin correlator contains contributions from multi-magnon scattering events. In the SM, we examine the role of two-magnon scattering that, to leading order in $1/S$, comes from the $S^z$ operator in the quantization frame. The two-magnon states form a continuum in momentum and energy leading to further scattering channels that increase the conductance (Fig.~\ref{fig:G_No_Impurity}). This contribution can, in principle, obscure the presence of band gaps in the single magnon band structure \cite{SM}.

The method is inherently energy resolution limited as the narrower the energy window the longer averaging times must be to overcome the temperature varying current noise. A rough benchmark is an $O(1)$ meV resolution at $4.2$ K for reasonable averaging times. Since $O(10-100)$ meV magnetic energy scales are not uncommon one may expect to resolve details of the local magnetic density of states in many 2D magnetic materials.

%%%%%%%%%%%%%%%%%%%%%%%%%%%%%%%%%%%%%%%%%%%%%%%%%%%%%%%%%%%%%%%%%%%%%%%%%%%%%%%%%%%%%%%%%%%
%%% SECTION: QPI
%%%%%%%%%%%%%%%%%%%%%%%%%%%%%%%%%%%%%%%%%%%%%%%%%%%%%%%%%%%%%%%%%%%%%%%%%%%%%%%%%%%%%%%%%%%

{\it Impurity Scattering and QPI for Magnons} $-$ In conventional QPI, the breaking of translational invariance in the vicinity of spatial disorder, leads to a tunneling conductance with a spatial dependence that is tied to the electronic band structure. Here, in order to uncover more detailed information about the magnetic excitations in 2D magnets, we propose systematically introducing disorder into the system.
In the following, we consider the case where single vacancies are introduced into an otherwise pristine lattice of magnetic ions though similar considerations hold for any type of point-like disorder such as interstitial magnetic ions. We distinguish between two main cases: unfrustrated and frustrated magnetic interactions. An example of the former is the square lattice antiferromagnet. Such models have the feature that the magnetic ground state is stable to the presence of impurities.  In contrast, magnetic frustration destabilizes the magnetic structure around impurity sites.

\begin{figure}[!ht]
    \centering
  \includegraphics[width=\columnwidth,clip,trim={1.5cm 0.5cm 1.5cm 0.5cm}]{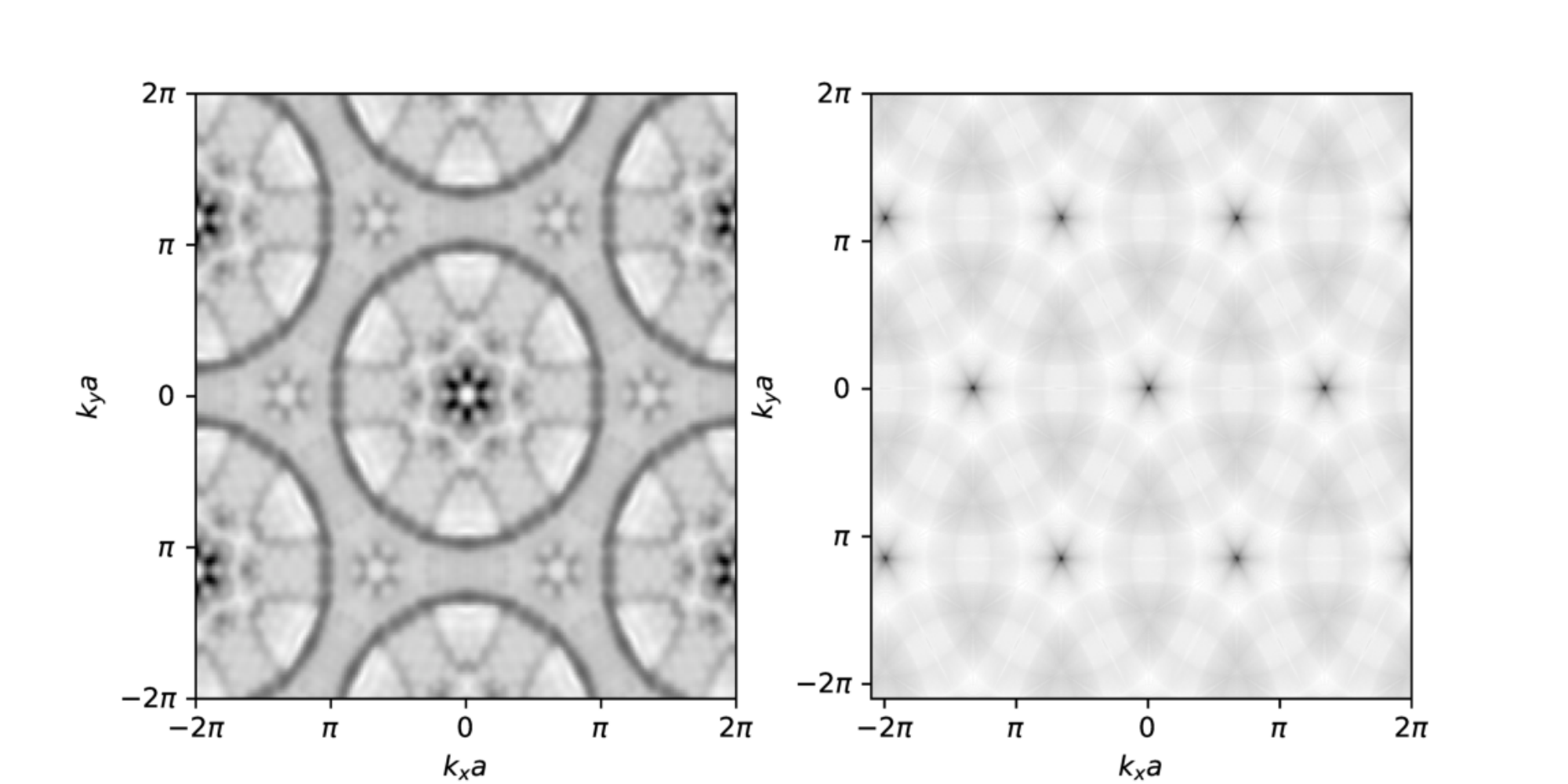}
        \caption{Magnon QPI results for the spin polarized Kitaev honeycomb model with a single vacancy and $\omega/KS=1.6$. The left-hand panel is the Fourier transform of the differential conductance, computed with $\delta V=0.1$, on the periodic 21x21 slab. The right-hand panel is the joint density of states. The bands extend from $1.0$ to $5.0$ in these units.}
    \label{fig:kitaev_qpi}
\end{figure}

We consider first the unfrustrated case and, in particular, the Neel ordered square lattice Heisenberg antiferromagnet. This model has linearly dispersing Goldstone modes at the zone centre and zone corners.  Because the moment orientation is undisturbed when a single magnetic site is removed, one may obtain the LDSF exactly. This can be done by writing the linear spin wave Hamiltonian as $H=H_0 + V$ where $H_0$ is the spin wave Hamiltonian for clean system and  $V$ is the impurity contribution. The single particle Matsubara Green's functions in terms of the Holstein-Primakoff bosons for the full problem $[\mathcal{G}(i\omega_n)]_{ij}$ satisfy a Dyson equation $[\mathcal{G}(i\omega_n)]_{ij} = [\mathcal{G}_0(i\omega_n)]_{ij} + [\mathcal{G}(i\omega_n)]_{im} V_{mn} [\mathcal{G}_0(i\omega_n)]_{nj}$ that can be solved numerically exactly based on analytical Green's functions $\mathcal{G}_0(i\omega_n)$ for the spin wave theory in the absence of impurities \cite{SM}. 
With $[\mathcal{G}(i\omega_n)]_{ii}$, one may evaluate the conductance in Eq.~(\ref{eq:conductance}) to leading order because for the problems we shall consider using this approach $\tilde{S}^{xx}_{ii}(\omega) = \tilde{S}^{yy}_{ii}(\omega) = -(1/\pi){\rm Im} [\mathcal{G}_{\rm ret}(\omega)]_{ii}$ where $\mathcal{G}_{\rm ret}(\omega)$ is the retarded single particle Green's function in terms of Holstein-Primakoff bosons \cite{SM}. 
Further assuming $\lambda/a\ll 1$, the differential conductance, $\partial_V G(V)$, becomes proportional to the LDSF.

In Fig.~\ref{fig:impurity}(left panel) the LDSF is plotted at fixed energy $\omega/4JS = 0.875$ where $4JS$ is the magnon bandwidth. The ripples extending out from the impurity correspond to the momentum space pattern in the middle panel. Further energy slices are shown in the SM \cite{SM}. In real space, as the energy increases from zero, roughly circular waves emanate from the impurity with decreasing wavelength as the voltage increases (see Fig.~$5$(a,b) in the SM). These have a direct interpretation in terms of the linearly dispersing Goldstone mode. In the corresponding Fourier transformed pattern the differential conductance has rings around the $(0,0)$ and $(\pi,\pi)$ points with radius $k_{\rm max} = 2\omega/v$ where $v$ is the velocity of the mode. At higher energies, the magnon band flattens out and more scattering channels open up \cite{SM} with the result that (i) the real space pattern evolves from circular waves at low energies to a $C_4$ symmetric pattern at higher energies and (ii)
the Fourier transformed pattern acquires much richer features at higher energies than at lower energies. Nevertheless these patterns have a direct interpretation in terms of the joint density of states $\rho(\bs{q},\omega)= \int d^2\bs{k}/(2\pi)^2 \delta(\omega - \epsilon_{\bs{q}+\bs{k}}) \delta(\omega-\epsilon_{\bs{k}})$ plotted in the rightmost panel. The resemblance between the Fourier transform of the constant energy local Green's functions and the joint density of states (JDOS) holds for all energies \cite{SM} highlighting the importance of the scattering of two magnons on a constant energy surface  $\epsilon_{\bs{k}}=\epsilon_{\bs{k}+\bs{q}}$  \cite{SM}. The band structure can, in principle, be inferred from experimental data by parametrizing the joint density of states.

Exact calculations of this kind can also be carried out for the two-band honeycomb lattice ferromagnet. In this model the two bands are connected by Dirac points at $K$ and $K'$ points. Results are shown in the SM \cite{SM}. 

\begin{figure}
  \centering
  \includegraphics[width=\columnwidth,trim=0.1cm 0.1cm 0.1cm 0.1cm,clip]{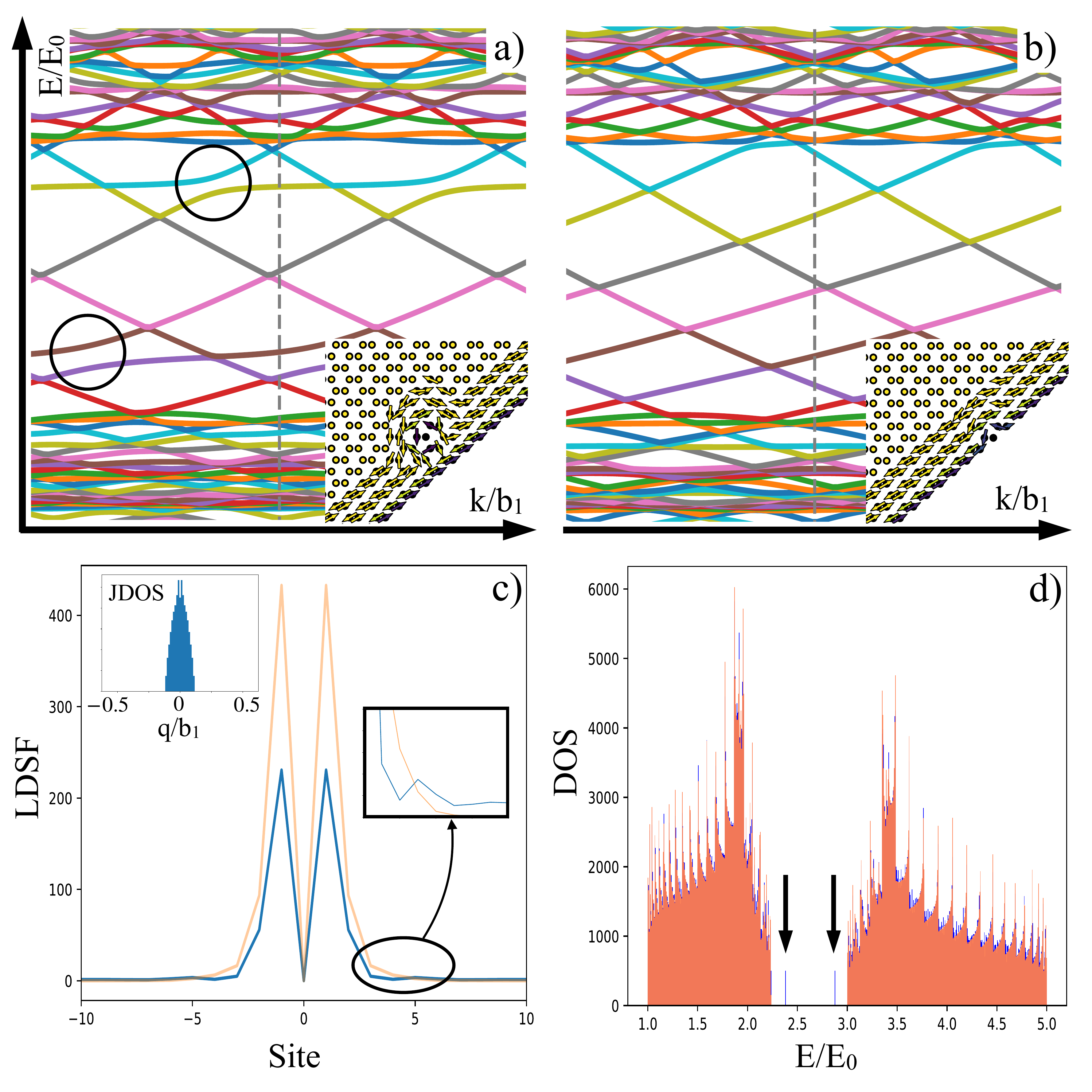}
  \caption{Calculations for honeycomb Kitaev-Heisenberg model
  with $\theta=\pi/2$ and magnetic field $h=5$ perpendicular to the honeycomb plane.
  a) Hybridized in-gap magnon states in the presence of a defect in an intermediate position close to the edge (full band dispersion in the SM \cite{SM}). 
  b) Renormalized states for a defect at the edge.
  For the above panels, insets show the respective ground states.
  c) Local dynamical structure factor for a case with a defect in the bulk (orange) and for a defect hybridizing with the boundary states (blue). The zoom shows the interference fringe and the inset shows the joint density of states (JDOS).
  d) Density of states (DOS) for the bulk without defect (pink) and with a defect in the bulk (blue). The two arrows highlight the localized bound states in the gap.
}
\label{fig:Topological_Bands}
\end{figure}

As a final example, we carry out a full calculation of the tunneling conductance on a finite periodic slab for the Kitaev-Heisenberg model in a magnetic field. Because the model is frustrated, the presence of an impurity destabilizes the polarized state locally. The ground state texture is one where the moments acquire transverse components that wind around the impurity. The sense of the winding depends on the distance from the impurity and the decay length scale of the texture is related to the inverse magnon gap that can be controlled by a uniform magnetic field (see Fig.~\ref{fig:Topological_Bands}(insets in a, b) \cite{SM}).

Despite the destabilization of the ground state and finite size effects, the differential tunneling conductance remains in close correspondence with the joint density of states computed from the bulk band structure (see Fig.~\ref{fig:kitaev_qpi} \cite{SM}).

%%%%%%%%%%%%%%%%%%%%%%%%%%%%%%%%%%%%%%%%%%%%%%%%%%%%%%%%%%%%%%%%%%%%%%%%%%%%%%%%%%%%%%%%%%%
%%% SECTION: TOPOLOGICAL MAGNON BANDS
%%%%%%%%%%%%%%%%%%%%%%%%%%%%%%%%%%%%%%%%%%%%%%%%%%%%%%%%%%%%%%%%%%%%%%%%%%%%%%%%%%%%%%%%%%%

{\it Topological Magnon Bands} $-$ The Kitaev-Heisenberg model on the honeycomb lattice in the polarized phase has two magnon bands carrying Chern number $\pm 1$ \cite{McClarty2018,Joshi2018,McClarty2021}. Nontrivial band topology of this type implies the presence of chiral magnon edge states. Detecting these edge states is a challenge for experiment because the modes are microscopic $-$ in the sense of an atomically small length scale binding them to the edge $-$ and also because the modes carry no charge and have energy scales significantly below electronic energy scales. 

Edge modes are detectable in principle as an in-gap contribution to the differential conductance when the tip is positioned at the edge \cite{Feldmeier2020} although there may be contamination from magnon continua or bound states. Since the system is translationally invariant parallel to a straight edge, the contribution to the tunneling coming from the edge modes is uniform in that direction. An important question is whether tunneling spectroscopy assisted by impurity scattering can be used to establish the chirality of in-gap modes or that they are topologically protected.

We return to the situation where a vacancy is introduced into the bulk focussing now on the in-gap region. One may loosely view the vacancy as a hole carrying protected chiral edge states that is shrunk down to atomic proportions. In this limit, the edge states on facing surfaces interact but, instead of destroying the edge modes completely, bound impurity states remain that within the bulk gap (Fig.~\ref{fig:Topological_Bands}(d)). Such localized in-gap states appear as jumps in the conductance at discrete in-gap voltages when the tip is placed close to the impurity. 
The presence of such bound states has been argued to be a diagnostic of topological bands at least when the hopping amplitudes are spatially homogeneous \cite{Slager2015,Diop2020}.

We now consider the QPI pattern found from the tunneling conductance around an impurity for different impurity positions with respect to a boundary. As discussed above, the QPI pattern in momentum space roughly reflects the joint density of states computed from the band structure in the translationally invariant system apart from the presence of the in-gap impurity bound states. 

Inspection of the spectrum of the model with impurity close to, but not at, the boundary in a slab calculation reveals that the nearly flat bound impurity modes hybridize with the edge modes leading to avoided crossings over narrow energy ranges (Fig.~\ref{fig:Topological_Bands}(a)). This leads to the following picture for the tunneling conductance as the voltage is swept through the in-gap region. For Chern bands supporting a single chiral edge mode, the joint density of states has a single peak at $\bs{q}=0$ so one should expect a spatially uniform, or trivial, QPI pattern in real space around an impurity except close to the hybridization energies {\it even though there are tunneling contributions originating from the edge modes}. Over the narrow energy window where hybridization occurs, scattering is possible between the admixed impurity and edge state modes leading to a broadening in the joint density of states that may be observed as a long wavelength spatial modulation in the conductance parallel to the edge (Fig.~\ref{fig:Topological_Bands}(c)). Altogether this provides evidence for the chirality of the edge modes.

If the vacancy lies directly at the boundary thereby locally roughening the edge, the topological nature of the bands implies that chiral surface states must be present. The principal difference between the case with and without the impurity is that the velocity of the chiral modes is renormalized (Fig.~\ref{fig:Topological_Bands}(d)). The robustness of the edge modes to the presence of disorder constitutes further evidence that the boundary modes are protected by the bulk topology.

\begin{acknowledgments}
We thank Jeff Rau and Masafumi Udagawa for useful discussions. A.M. acknowledges the support of the KVPY fellowship from DST, Govt. of India. A.C. and P.M. acknowledge the Deutsche Forschungsgemeinschaft (DFG) Grant No. SFB 1143. 
P.R. acknowledges support by FCT through Grant No. UID/CTM/04540/2019.
\end{acknowledgments}

\bibliography{references}

\clearpage

\addtolength{\oddsidemargin}{-0.75in}
\addtolength{\evensidemargin}{-0.75in}
\addtolength{\topmargin}{-0.725in}

\newcommand{\addpage}[1] {
 \begin{figure*}
   \includegraphics[width=8.5in,page=#1]{Magnon_QPI_SM.pdf}
 \end{figure*}
}
\addpage{1}
\addpage{2}
\addpage{3}
\addpage{4}
\addpage{5}
\addpage{6}
\addpage{7}
\addpage{8}
\addpage{9}
\addpage{10}
\addpage{11}
\addpage{12}
\addpage{13}
\addpage{14}
\addpage{15}
\addpage{16}
\addpage{17}
\addpage{18}
\addpage{19}
\addpage{20}

\end{document}